\documentclass[aps,prd,groupedaddress,showpacs,graphicx,nofootinbib]{revtex4}
\usepackage{amssymb,graphics,graphicx,color}

\begin{document}

\title{Hilltop Supernatural Inflation and SUSY Unified Models}

\author{Kazunori Kohri$^{1}$}\email{kohri@post.kek.jp}
\author{C. S. Lim$^{2}$}\email{lim@lab.twcu.ac.jp}
\author{Chia-Min Lin$^{3}$}\email{lin@chuo-u.ac.jp}
\author{Yukihiro Mimura$^{4}$}\email{mimura@hep1.phys.ntu.edu.tw}

\affiliation{$^1$Cosmophysics group, Theory Center, IPNS, KEK}
\affiliation{$^2$Department of Mathematics, Tokyo Woman's Christian University, Tokyo 167-8585, Japan}
\affiliation{$^3$Department of Physics, Chuo University, Bunkyo-ku, Tokyo 112, Japan}
\affiliation{$^4$Department of Physics, National Taiwan University, Taipei, 10617, Taiwan (R.O.C.)}

\date{Draft \today}

\begin{abstract}
In this paper, we consider high scale ($100$ TeV) supersymmetry (SUSY) breaking and realize the idea of hilltop supernatural inflation in concrete particle physics models based on flipped-SU(5) and Pati-Salam models in the framework of supersymmetric grand unified theories (SUSY GUTs). The inflaton can be a flat direction including right-handed sneutrino and the waterfall field is a GUT Higgs. The spectral index is $n_s=0.96$ which fits very well with recent data by PLANCK satellite. There is no both thermal and non-thermal gravitino problems. Non-thermal leptogenesis can be resulted from the decay of right-handed sneutrino which plays (part of) the role of inflaton. 
\end{abstract}
\maketitle

\section{Introduction}

\large
\baselineskip 18pt

Inflation \cite{Lyth:2009zz} is becoming a standard model for the very early universe. However, there is still no consensus about questions like: What is the inflaton field? How do we connect inflation to particle physics? How does reheating connect to standard model? There are lots of inflation models and many of them simply cannot address on these questions. Among the inflation models, \emph{hybrid inflation} is a promising one in order to connect inflationary cosmology to particle physics. In the framework of SUSY, there are $F$- and $D$-term hybrid inflation models. However in those models the inflaton is usually assumed to be some unknown gauge singlet and it is not clear how does inflaton decay and reheat the universe to result in a thermal bath of standard model particles in order to have successful big bang nucleosynthesis. Furthermore, these models may suffer severely from thermal \cite{Kawasaki:2004qu} and non-thermal gravitino problem \cite{Kawasaki:2006gs}. On the other hand, there is still another interesting SUSY hybrid inflation model which is called \emph{supernatural inflation} \cite{Randall:1995dj}. However, the original form of this model predicts a blue spectrum, namely the spectral index $n_s>1$. Current observation from PLANCK satellite \cite{Ade:2013ydc} gives $n_s = 0.9603 \pm 0.0073$ which excludes $n_s=1$ at over $5\sigma$. PLANCK also reports smaller tensor to scalar ratio $r<0.11$ and no non-Gaussianity is observed. This implies single-field slow-roll inflation is still a very good model as long as the spectrum is red. If we further restrict ourselves to small field inflation with inflaton field value smaller than Planck scale, it is almost inevitable to consider hilltop inflation \cite{Boubekeur:2005zm} where the potential is convex (corresponding to tachyonic fluctuation modes), namely the slow-roll parameter $\eta<0$ at horizon exit.  In \cite{Lin:2009yt}, techniques of type-III hilltop inflation developed in \cite{Kohri:2007gq} is applied to supernatural inflation and obtain the spectral index $n_s=0.96$ with natural scales coming from SUSY breaking. It was later shown in \cite{Kohri:2010sj} (see also \cite{Lin:2010zzk}) that this \emph{hilltop supernatural inflation} can evade both thermal and non-thermal gravitino problems. In this paper, we materialize this model in some solid SUSY GUT models including flipped-SU(5) and Pati-Salam model\footnote{Hybrid inflation in the framework of flipped-SU(5) and Pati-Salam models was considered in \cite{Kyae:2005nv, Jeannerot:2000sv, Antusch:2010va}. However, these papers are based on $F$-term hybrid inflation (with modification). On the other hand, we consider (hilltop) supernatural inflation in this paper.}.

This paper is organized as follows: In section \ref{sec1}, we briefly review the model of hilltop supernatural inflation. In section \ref{section3}, we present the potential. In section \ref{section4}, flipped SU(5) model is considered. In section \ref{section5}, Pati-Salam model is considered. 
In section \ref{section6}, we describe the topological defects after the inflation in our models.
We discuss issues after inflation including gravitino problem and leptogenesis in section \ref{section7} and section \ref{con} is our conclusion. In Appendix \ref{a}, we present some calculation detail for hilltop inflation models and in Appendix \ref{b}, we review flipped SU(5) and Pati-Salam models.

\section{Hilltop Supernatural Inflation}
\label{sec1}
The potential for a hybrid inflation is given by
\begin{equation}
V=\frac{1}{2}m_\psi^2 \psi^2 +g^2 \psi^2 \phi^2 + \kappa^2 (\phi^2-\Lambda^2)^2,
\label{eq1}
\end{equation}
where $\psi$ is the inflaton field and $\phi$ is the waterfall field. 
The effective mass of the waterfall field (at $\phi=0$) is
\begin{equation}
m^2_{\phi} \equiv V^{\prime\prime}(\phi)=2g^2 \psi^2 -4\kappa^2 \Lambda^2.
\label{eq2}
\end{equation}
During inflation, the field value of $\psi$ gives a large positive mass to $\phi$ therefore it is trapped to $\phi=0$ and the potential 
during inflation is of the form 
\begin{equation}
V=V_0+\frac{1}{2}m_\psi^2 \psi^2,
\label{eq3}
\end{equation}
where $V_0=\kappa^2 \Lambda^4$.
The end of inflation is determined by $m^2_{\phi}=0$ when the waterfall field starts to become tachyonic which implies
\begin{equation}
\psi_{end}=\frac{\sqrt{2}\Lambda \kappa}{g}.
\label{eq4}
\end{equation}

\begin{figure}[t]
  \centering
\includegraphics[width=0.6\textwidth]{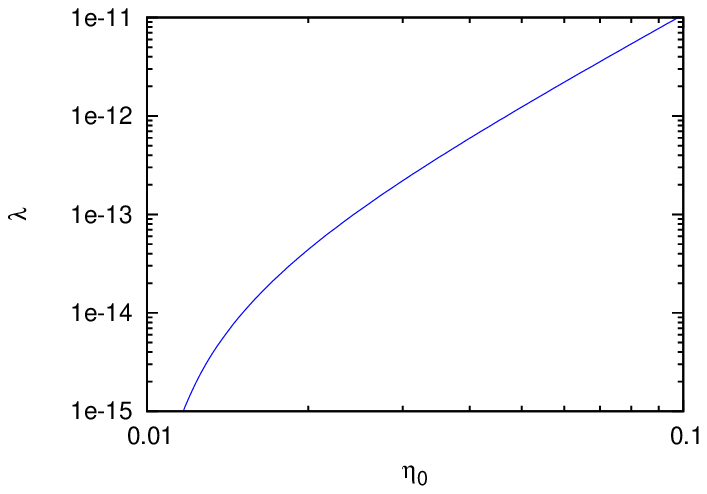}
  \caption{$\lambda$ as a function of $\eta_0$ for $p=4$. This plot is made by imposing $n_s=0.96$ and $P_\zeta^{1/2}=5 \times 10^{-5}$.}
  \label{fig1}
\end{figure}

For original supernatural inflation, because the potential in Eq.~(\ref{eq3}) is concave upward, 
a blue spectral index $n_s>1$ is obtained in the simplest form of this model. 
We can get a red spectral index if we extend the model into a hilltop supernatural inflation 
by considering 
\begin{equation}
V(\psi)=V_0+\frac{1}{2}m_\psi^2 \psi^2 - \lambda \psi^4 \equiv V_0\left( 1+\frac{1}{2}\eta_0 \frac{\psi^2}{M_P^2} \right) - \lambda \psi^4,
\label{main}
\end{equation} 
where $\eta_0 \equiv m_\psi^2 M_P^2/V_0$ and
$M_P=2.4 \times 10^{18}\ \mbox{GeV}$ is the reduced Planck mass. 
Given the potential, we can solve for the spectrum, spectral index, and the field value. 
We put the detailed calculation in the Appendix \ref{a}. 
The potential becomes concave downward (for $\lambda >0$) 
when cosmological scales leave the horizon at $N=60$ 
and as can be seen in Fig~{\ref{fig1}} a spectral index $n_s=0.96$ can be obtained by 
$\eta_0 = 0.02$ (for $\lambda = 4.4\times 10^{-14}$),
$0.03$ (for $\lambda = 2.2\times 10^{-13}$).

The quartic term in the scalar potential with a tiny coupling constant can be obtained
by considering a non-renormalizable term in the superpotential\footnote{We consider $\psi$ field as a flat-direction, therefore no renormalizable term are relevant here.}: 
\begin{equation}
W=a \frac{\psi^4}{M_P},
\end{equation}
where $a$ is a dimensionless coupling constant. 
This makes the quartic term in the scalar potential during inflation\footnote{There is also a positive $F$-term $\sim \psi^6$ but we can easily check that it is negligible due to the smallness of the field value $\psi$.} 
and $\lambda \equiv aA/M_P$, where $A$ is a SUSY breaking mass parameter.
If we choose a high-scale SUSY breaking $m_{\rm SUSY} \sim 100$ TeV,
this makes the effective coupling to be $\lambda \sim 10^{-13}$ for $a \sim O(1)$.
This interesting way to make a small coupling naturally is one of the reason our model can work. 
To obtain a suitable size of the coupling, high-scale SUSY breaking ($m_{\rm SUSY} = O(10)-O(100)$ TeV)
is well-motivated.


Different from the original supernatural inflation, one can consider $100$ TeV SUSY breaking 
which can be realized in a hidden sector model
in which $V_0=M_S^4$ 
where $M_S = \sqrt{m_{\rm SUSY} M_P} \sim 5\times 10^{11}\ \mbox{GeV}$ is the 
gravity-mediated SUSY breaking scale and $m_\psi \sim O(100)$ TeV is the soft mass.
The quantity $\eta_0=0.02$ can be achieved by $m_\psi = 14 \mbox{ TeV}$. 
Therefore roughly one order of adjustment of $m_\psi$ is required, though it is not fine-tuning.

In this paper, we address to construct the hilltop supernatural hybrid inflation 
in unified models.
In the models, $V_0 \equiv \kappa^2 \Lambda^4$ is a parameter in the GUT potential,
which is not necessarily related to the SUSY breaking.
Let us describe model-independent constraints and features of the type-III hilltop inflation.
\begin{enumerate}
\item
As can be seen in Fig.~\ref{fig2} and \ref{fig3}, 
in this model $\psi(N=60) \sim \psi_{end} \sim 10^{-7}M_P$ is obtained. 
This also implies a small tensor-to-scalar ratio and primordial gravity waves is unobservable.
\item
There is a lower bound for $\Lambda$ in our parametrization which can be obtained by imposing $g<1$ 
to avoid the model becoming non-perturbative. 
Supposing $V_0=\kappa^2 \Lambda^4 \sim (10^{12}\ \mbox{GeV})^4\sim 10^{-24}M_P^4$, 
we can write $\kappa \sim 10^{-12}M_P^2/\Lambda^2$. By using Eq.~(\ref{eq4}) with $\psi_{end} \sim 10^{-7}M_P$, we obtain $\Lambda g=10^{-5}M_P$. Therefore $g<1$ implies $\Lambda>10^{-5}M_P$.
\item
In order to make inflation ends promptly once the waterfall field becomes tachyonic, 
we require $|m_\phi^2| \gg H^2$ when $\psi$ approaches to the origin. 
By using Eqs.~(\ref{eq1}) and (\ref{eq2}), 
we have $|m_\phi^2|\sim \kappa^2 \Lambda^2$ and $H^2 \sim V_0/M_P^2=\kappa^2 \Lambda^4/M_P^2$. 
Therefore the condition  $|m_\phi^2| \gg H^2$ implies $\Lambda^2 \ll  M_P^2$. 
As we will see in the following sections, this condition is automatically satisfied.
\end{enumerate}

\section{The potential}  
\label{section3}
In this section, we present the potential form which will be used in subsequent sections for concrete particle physics models in the framework of SUSY GUT.

%
%

The scalar potential in supergravity is given by
\begin{equation}
V = e^{\frac{K}{M_P^2}} \left( K^{ij^*} F_i F_{j^*} - 3 \frac{|W|^2}{M_P^2}\right), 
\end{equation}
where
\begin{equation}
F_i = D_i W \equiv \frac{\partial W}{\partial \phi_i} + K_i W \frac{1}{M_P^2}, 
\qquad K_i = \frac{\partial K}{\partial \phi_i}.
\end{equation}
The gravitino mass is
\begin{equation}
m_{3/2} = e^{\frac{K}{2M_P^2}} \frac{W}{M_P^2}.
\end{equation}
Using the condition of elimination of cosmological constant 
$K^{ij^*} F_i F_{j^*} (\equiv |F|^2) = 3 \frac{|W|^2}{M_P^2}$,
we usually express the gravitino mass as
\begin{equation}
m_{3/2} = e^{\frac{K}{2M_P^2}} \frac{|F|}{\sqrt3 M_P}.
\end{equation}


We consider the following superpotential.
\begin{equation}
W (\phi,\psi) = W_0 + W(\phi) + a \frac{\psi^4}{M_P} + 
g \phi \psi \psi^\prime.
\label{eq11}
\end{equation}
We note that the $\psi^4$ term does not necessarily consist of a single $\psi$ field,
but it can be given by multiple fields as $\psi^4 = \psi_1 \psi_2 \psi_3 \psi_4$ 
satisfying 
$|\psi_1| = |\psi_2| = |\psi_3| = |\psi_4|$, which we will describe later.

The hilltop potential (for $\phi =0$) is
\begin{equation}
V(\phi=0,\psi) = e^{K/M_P^2}
\left(\sum_i
\left|a \frac{\prod_{j\neq i}\psi_j}{M_P} + \psi^*_i 
\left(W_0 + a \frac{\psi^4}{M_P}\right)\frac1{M_P^2}\right|^2
+ F_X F^X
-3 \left|W_0 + a \frac{\psi^4}{M_P}\right|^2\frac{1}{M_P^2}
\right)+V_0,
\end{equation}
where the K\"ahler potential for the matter field is assumed to be canonical,
and $F_X$ is an $F$-term of SUSY breaking superfield $X$. We assume $\psi^\prime=0$ which is the case if it is heavy during inflation. As we will see, for example, the field $N$ in Eq.~(\ref{eq46}) can play the role of $\psi^\prime$. The term $V_0$ is from $W(\phi)$ which will be addressed in Eq.~(\ref{eq15}).
If the condition $W_0 \gg a \psi^4/M_P$ is satisfied\footnote{
At the hilltop, $\psi^2 W_0^2/M_P^2 \sim a W_0 \psi^4/M_P$.
Therefore, the condition $W_0 \gg a \psi^4/M_P$
can be satisfied at the top of the potential.
},
the potential can be ``hilltop",
and we obtain
\begin{equation}
V (\psi) = V_0 + m_{3/2}^2 |\psi_i|^2 + \left(a A \frac{\psi^4}{M_P} + c.c \right) + \cdots,
\label{eq13}
\end{equation}
where $A = m_{3/2} e^{K/2M_P^2}(1+W_X K^X/W_0) \sim m_{3/2}$.
%
One can show that $\psi^6$ term (and higher order terms) 
can be negligible around the hilltop of the potential.
We note that 
the potential can be written as
\begin{equation}
V (\psi) = V_0 + m_{3/2}^2 |\psi_i|^2 + \lambda |\psi|^4 \cos\theta,
\end{equation}
where $\theta$ is a phase of $aA\psi^4$ and $\lambda = 2 |a A|/M_P$.
The hilltop configuration can be obtained
if the inflation starts at $\theta = \pi$.

The term $V_0$ can be obtained from the superpotential:
\begin{equation}
W(\phi) = \kappa S ( \bar{\phi}\phi - \Lambda^2).
\label{eq15}
\end{equation}
The key feature of this potential is that there is only linear term of a singlet field $S$,
and the mass term of $S$ and cubic term are forbidden (or suppressed by an approximate symmetry) \footnote{However, the field $S$ is heavy with a mass roughly the inflation scale due to a coupling to the squark condensation (in addition to the waterfall field) which can generate the required inflation scale. Therefore $S$ cannot be the inflaton in our model. See the following discussion around Eq.~(\ref{eq17}).}.
This can be achieved by $R$-symmetry.

Combining  Eqs.~(\ref{eq11}), (\ref{eq13}) and (\ref{eq15}) and setting $S=\psi^\prime=0$, the whole potential in our model is given by
\begin{equation}
V(\phi,\psi)=m_\psi^2|\psi|^2+aA\frac{\psi^4}{M_P}+|g\psi\phi|^2+|\kappa(\bar{\phi}\phi-\Lambda^2)|^2.
\end{equation}
This can be compared with Eq.~(\ref{eq1}) (by rotating the phase to the real component of the complex scalar fields) except the second term which is introduced to make the potential into a hilltop form by choosing a negative $a$. During inflation, the large expectation value of the inflaton field $\psi$ would force $\phi=0$ through the third term in the potential therefore the fourth term becomes $V_0=\kappa^2 \Lambda^4$ which provides the vacuum energy to drive inflation.

We note that the linear term of the singlet field $S$ 
can be always generated due to the SUSY breaking \cite{Bagger:1993ji}.
In fact, the following term in the K\"ahler potential 
can be always generate the linear term
\begin{equation}
\int d^4 \theta \frac1{M_P} X X^\dagger S = \int d^2\theta \frac{1}{M_P}X F_X^\dagger S,
\end{equation}
where $X$ is a SUSY breaking superfield, whose $F$-term is non-zero.
Because the gravitino mass is $m_{3/2} \simeq F_X/M_P$,
the size of $\kappa \Lambda^2 = V_0^{1/2}$ is related to the SUSY breaking $m_{3/2} X \sim m_{3/2} M_P$.


%

%
The scale $\kappa\Lambda^2$ can be also obtained by strong $SU(N)$ dynamics
(which is different form color $SU(3)_c$).
The condensation of ``squark" fields of $SU(N)$, $\langle \bar Q Q \rangle$, 
can induce the dynamical scale $\Lambda$
via $S \bar Q Q$ interaction term (In this example of model, thus, $\Lambda$ has non-trivial $R$-charge),
e.g.
\begin{equation}
W = S \bar Q Q + \bar\mu (\det M  - B\bar B - \bar\Lambda^{2N}),
\label{eq17}
\end{equation}
where $M$, $B$ and $\bar B$ are ``meson" and ``(anti)baryon" condensations
and $\bar\mu$ is a Lagrange multiplier \cite{Seiberg:1994bp}.
In this building of waterfall potential,
the condensation $\langle \bar Q Q \rangle$ generates
$V_0^{1/2} = \bar\Lambda = \kappa \Lambda^2$,
which can be much less than the Planck scale,
and it is a free parameter in the model.
Smallness of the coupling constant $\kappa$ induces
the hierarchy between $V_0^{1/4}$ and the VEV of the waterfall field (namely, unification scale).
We do not address the detail of the mechanism in this paper,
and we mention the feature of this potential.



It is well-known that fine-tuning between the 
SUSY breaking order parameter $F_X$ and superpotential $W$
is needed to eliminate the cosmological constant
\begin{equation}
K^{ij^*} F_i F_{j^*} = 3 \frac{|W|^2}{M_P^2}.
\end{equation}
%
In order to realize the proper SUSY breaking scale, we need
\begin{eqnarray}
&&F^{1/2} \sim 10^{10} - 10^{11} \ {\rm GeV}, \\
&&W^{1/3} \sim 10^{13} \ {\rm GeV}.
\end{eqnarray}

In naive GUT superpotential $W_{\rm GUT}$,
one obtains $W_{\rm GUT}^{1/3} \equiv M_G \sim 10^{16}$ GeV.
We need $3|\hat W+M_G^3|^2/M_P^2 =|F|^2$ ($\hat W$ is a non-GUT superpotential) 
for vanishing the cosmological constant.
Therefore, to realized the proper SUSY breaking scale and vanishing the cosmological constant, 
two-step cancellation is needed (among three quantities):
\begin{equation}
|\hat W + M_G^3|- M_P |F| \ll M_P |F| \simeq |\hat W + M_G^3| \ll M_G^3,
\end{equation}
unless no-scale supergravity is considered.
If the waterfall potential $V(\phi)$ is employed to break GUT symmetry,
the first cancellation in the superpotential is ``automatic"
via the $F$-flatness condition.
(Surely, we still need a fine-tune between SUSY breaking and 
total $W$ for vanishing cosmological constant).
The waterfall superpotential requires only one cancellation
even if a field acquires a VEV $\sim 10^{16}$ GeV.  
This is one of the important conceptual merits of this scenario
in the view of GUT model building.
The size of $\langle W \rangle$ is related to the $R$-symmetry breaking scale,
which should related to the SUSY breaking order parameters.
In the supernatural inflation models, the size of $V_0^{1/4}$ can be also related to the
scale naturally.

Another implication of the waterfall potential comes from a symmetry
of the potential.
The symmetry can be utilized to suppress dangerous proton decay operators.
Due to the symmetry,
some of the fields remains light ($\sim \kappa \Lambda$) in the multiplet
to break GUT symmetry.
Contrary to the usual GUT superpotential, cubic terms of the field whose VEV breaks GUT symmetry
is absent.
As a consequence, there can be an accidental global symmetry (which may be softly broken),
and the fields can be light to be TeV scale (or SUSY breaking scale),
which can have an phenomenological implication.

\section{Unified Model Building}
\label{section4}

In the potential of the hybrid inflation, 
a large VEV of the waterfall field is suggested.
The large VEV is available to break a unified symmetry.
As we have mentioned in the previous section,
the waterfall potential should have a symmetry
which can be adopted to explain a phenomenological issue in the unified models.
To explain these features,
we consider the unified model, in which the unified gauge symmetry is
directly broken down to SM by a VEV of the waterfall field,
(1). flipped-SU(5) model \cite{DeRujula:1980qc,Nanopoulos:2002qk}, 
whose gauge symmetry is $SU(5) \times U(1)_X$, 
(2). Pati-Salam model \cite{Pati:1974yy}, 
whose gauge symmetry is $SU(4)_c \times SU(2)_L \times SU(2)_R$.
The brief introductions of the models are given in Appendix \ref{b}.

\subsection{Waterfall potential}

%
%
In the flipped-SU(5) model,
the gauge symmetry
can be broken by
\begin{eqnarray}
T : ({\bf 10}, 1),
\quad
\bar T : (\overline{\bf 10}, -1),
\end{eqnarray}
under the $SU(5) \times U(1)_X$ symmetry.
The {\bf 10} representation under SU(5) is two-rank anti-symmetric tensor.
The VEVs of $T_{45}$ and $\bar T^{45}$
(those are singlet components under SM gauge group)
can directly break $SU(5)\times U(1)_X$ down to SM.
The symmetry breaking can occur via the waterfall superpotential:
\begin{equation}
W = \kappa S ( T \bar T - \Lambda^2).
\end{equation}

As many people concern, GUT models 
have doublet-triplet splitting problem.
In the flipped-SU(5) model, the doublet-triplet splitting can be realized simply.
The Higgs multiplets which include MSSM Higgs doublets ($H_u, H_d$) are:
\begin{eqnarray}
H : ({\bf 5}, -2),
\quad
\bar H : (\bar{\bf 5}, 2).
\end{eqnarray}
%
%
%
The multiplets also include colored-triplets $H_C, \bar H_C$.
Then, the following superpotential is allowed.
\begin{equation}
W=TTH+\bar T \bar T \bar H.
\end{equation}
The VEV of $T^{45}$ and $\bar T^{45}$ ($= \langle T \rangle$) 
to break the GUT symmetry 
makes the colored triplet heavy
because $T$ multiplet has a colored-triplet Higgs component $T_C$ (so-called missing partner).
The Higgs doublet, on the other hand, does not acquire masses from $\langle T \rangle$.

The Higgs mass term is
\begin{equation}
W_H = \left(
\begin{array}{cc}
 T_C & H_C
\end{array}
\right)
M_{H_C}
\left(
\begin{array}{c}
 \bar T_C \\ \bar H_C
\end{array}
\right)
+ M_H H_u H_d,
\qquad
M_{H_C} = \left(
	\begin{array}{cc}
		M_T & \langle T \rangle \\
		\langle \bar T \rangle & M_H
	\end{array}
\right),
\end{equation}
where $M_T$ and $M_H$ are mass parameters of $T$ and $H$ multiplets. 
The doublet Higgs mass (usually called Higgsino mass $\mu$) should be small,
and therefore $M_H$ should be small,
which can originate from the accidental discrete symmetry (by assuming $R$-symmetry\footnote{The terms (Mass of $H\bar{H}$, $T\bar{T}$, $H\bar{H}T\bar{T}$, etc) which can spoil the DT splitting can be forbidden by R-symmetry.  The concrete R symmetry is written in \cite{Kyae:2005nv}.})
to construct the waterfall potential in GUT. 

The interesting feature of this mechanism is that
the proton decay via the exchange of colored Higgs fields (dimension 5 operator)
can be suppressed.
Indeed, we assume that the matter fields does not couple to $T$,
but couple to $H$ via Yukawa interaction (neglecting to show coupling constants 
and generation indices explicitly)
by a symmetry to construct the waterfall potential:
\begin{equation}
W_Y = \frac12 qq H_C + u^c e^c H_C + q\ell \bar H_C + d^c u^c \bar H_C.
\end{equation}
Then, integrating out the heavy colored Higgs field,
we obtain
\begin{equation}
W = -  \left(\frac12 qq + u^c e^c\right) ( q\ell  + d^c u^c) (M_{H_C}^{-1})_{22} ,
\end{equation}
and the operator is suppressed by a factor $M_T/\langle T\rangle$:
\begin{equation}
(M_{H_C}^{-1})_{22}
= \frac{M_T}{M_T M_H - \langle T\rangle\langle \bar T\rangle}
\sim
\frac{M_T}{\langle T\rangle} \frac1{M_G}.
\end{equation}

It is interesting to note that
$M_T$ has to be much smaller than
the GUT scale $\sim \langle T\rangle = \Lambda$
in our inflation framework.
In fact, even if we consider the correction from SUSY breaking sector,
the mass parameter $M_T$ is up to the SUSY breaking order parameter, $(m_{\rm SUSY} M_P)^{1/2}$.
This is responsible for a symmetry to obtain the waterfall potential for the hybrid inflation.
This means our inflation model is compatible with proton decay suppressing.

The inflation model in the flipped-SU(5) model with the waterfall potential
is also constructed in Ref.\cite{Kyae:2005nv}.
In their model, the singlet field is the inflaton.
In this paper, we construct the quartic hilltop potential term
in terms of the matter fields.

\subsection{Hilltop potential}

%

The hilltop potential can be constructed in terms of the matter superfields,
and the scalar partners of the quarks and leptons can cause the inflation.
Let us describe the hilltop potential in the flipped-SU(5) model.

The matter content of the flipped-SU(5) model (See Appendix \ref{b} for more detail) is 
\begin{equation}
\begin{array}{|c|c|}\hline
\mbox{Matter} &  SU(5) \times U(1)_X \\ \hline\hline
{\bf 10}_i = (q_i,d^c_i,\nu^c_i) & ({\bf 10},1) \\ \hline
\bar{\bf 5}_i = (u^c_i,\ell_i)& (\bar{\bf 5},-3)  \\ \hline
{\bf 1}_i = e^c_i & ({\bf 1},5) \\ \hline
\end{array}
\end{equation}
where $i$ is a generation index.
The Yukawa couplings are:
\begin{equation}
Y_u^{ij} {\bf 10}_i {\bar {\bf 5}}_j \bar H
+
Y_d^{ij} {\bf 10}_i {\bf 10}_j H
+
Y_\ell^{ij} {\bar {\bf 5}}_i {\bf 1}_j H.
\end{equation}
The Dirac neutrino Yukawa coupling is unified with the up-type quark Yukawa coupling $Y_u$.


Because the representation of 
the fields $T$ to break the $SU(5)\times U(1)_X$ symmetry down to SM
are same as the matter filed ${\bf 10}_i$,
one has to adopt a discrete symmetry (or $R$-symmetry) to distinguish them.
The symmetry can be also useful to obtain the waterfall potential.
Cubic terms of the matter representations are not invariant under the gauge symmetry.
In the usual SU(5) GUT, $\lambda_{ijk} {\bf 10}_i \bar{\bf 5}_j \bar{\bf5}_k$ term is allowed.
This term is not singlet under $U(1)_X$ in the flipped-SU(5) model.

The possible quartic terms in terms of the matter multiplets are
\begin{equation}
{\bf 10} \cdot {\bf 10} \cdot {\bf 10} \cdot \bar{\bf 5},
\qquad
{\bf 10} \cdot \bar{\bf 5} \cdot \bar{\bf 5} \cdot {\bf 1}.
\end{equation}
In usual SU(5) GUT, since $\bar{\bf 5}$ and $\bar H$ have same quantum charges,
and they can be replace.
However, in the flipped-SU(5) model, since they have different $U(1)_X$ charges, the replacement is forbidden.
We note that the quartic Higgs term $H H \bar H \bar H$ is allowed.
It can however destabilize the electroweak symmetry breaking (which is the same circumstances
as in MSSM) due to a diagram with quadratic divergence,
and both $H \bar H$ and $H H \bar H \bar H$ have to be suppressed by a symmetry

The auxiliary field in the vector multiplet (so-called $D$-term) generates the $D$-term potential.
The K\"ahler potential of the matter superfields
are given as
\begin{equation}
K = \frac12 {\bf 10}^\dagger{}^{ab} {\bf 10}_{ab} + \bar{\bf5}^\dagger_a \bar{\bf5}^a + {\bf1}^\dagger {\bf1}.
\end{equation}
Then, the $D$-term for $SU(5)$ and $U(1)_X$ are obtained as
\begin{eqnarray}
D_5^A &=& (T^A)^a_b ({\bf 10}_{ac} {\bf 10}^\dagger{}^{bc} - \bar{\bf5}^\dagger_a \bar{\bf5}^b) \equiv 
(T^A)^a_b (D_5)^b_a \\
D_1 &=& \frac{1}{2}{\bf 10}_{ab} {\bf 10}^\dagger{}^{ab} -3 ( \bar{\bf5}^\dagger_a \bar{\bf5}^a )+ 5 ({\bf1}^\dagger {\bf1}).
\end{eqnarray}


One can find that $D$-flatness (vanishing $D$-term potential) condition
is satisfied, for example, for
\begin{equation}
|\psi| = |{\bf 10}^{\it 12}_1| = |\bar{\bf 5}^{\it 1}_1| = |\bar{\bf 5}^{\it2}_2| = |{\bf 1}_1|, 
\label{Dflat-fSU5}
\end{equation}
where subscripts denote the flavor indices
and the italic superscripts denote the gauge group indices.
We stress that
the $F$-term potential is lifted-up
if the generation indices are given as their mass eigenstates:
\begin{equation}
\left|\frac{\partial W}{\partial \bar H^{\it2}}\right|^2
= |Y_u^{11} {\bf 10}_{1}^{\it 12} \bar{\bf 5}_1^{\it1} + \cdots |^2.
\end{equation}
The up-quark Yukawa coupling is about $10^{-5}$,
and the coupling of quartic term is about $10^{-10}$, which is not negligible.
The $F$-term potential always provides a concave upward piece,
which is not suitable to obtain $n_s < 1$. 
Without loss of generality,
using the unitary rotation in the generation space,
one can take a basis where
\begin{equation}
Y_u = \left(
\begin{array}{ccc}
0 & 0 & \mbox{x} \\
0 & \mbox{x} &  \mbox{x}\\
 \mbox{x} & \mbox{x}& \mbox{x}
\end{array}
\right), \quad
Y_d = \left(
\begin{array}{ccc}
0 & 0 & \mbox{x} \\
0 & \mbox{x} &  \mbox{x}\\
 \mbox{x} & \mbox{x}& \mbox{x}
\end{array}
\right), \quad
Y_\ell = \left(
\begin{array}{ccc}
0 & 0 & \mbox{x} \\
0 & \mbox{x} &  \mbox{x}\\
 \mbox{x} & \mbox{x}& \mbox{x}
\end{array}
\right),
\end{equation}
where $\mbox{x}$ stands for any non-zero values.
Under this generation configuration,
both $F$- and $D$-flatness can be satisfied.
This is because there are three generations.
The components correspond to the SM component fields,
\begin{equation}
|\psi| = |d^{c}_b| = |u^{c}_r| = |c^{c}_g| = |e^c|.
\end{equation}
We are using the conventional flavor notation for simplicity, but these are not given as mass eigenstates.
Considering a non-renormalizable superpotential term
\begin{equation}
W = a \frac{1}{M_P} {\bf 10}_1 \bar{\bf 5}_1 \bar{\bf 5}_2 {\bf 1}_1,
\end{equation}
we obtain the hilltop potential term (up to a phase)
\begin{equation}
\frac{aA}{M_P}  \psi^4 = \frac{aA}{M_P} d^c u^c c^c e^c.
\end{equation}

One can also find that the $D$-flatness condition is satisfied for
\begin{equation}
|\psi| = |{\bf 10}^{\it 45}_1| = |{\bf 10}^{\it 12}_1| = |{\bf 10}^{\it23}_2| = |\bar{\bf 5}^{\it 2}_1|, 
\end{equation}
which corresponds to
\begin{equation}
|\psi| = |\nu^c| = |d^c_b| = |s^c_r| = |u^c_g|, 
\end{equation}
and
$\psi^4 = \nu^c d^c s^c u^c$.
We note that the $D$-term vanishes if $(D_5)_a^b \propto \delta_a^b$.
This is because the group generator is traceless in simple groups.
%
%
Through this superpotential term, baryon number $B$ and lepton number $L$
can be generated by Affleck-Dine (AD) mechanism, when the inflaton oscillates. 
However, $B-L$ is not generated.

Other example of the $D$-flat direction is 
\begin{equation}
|\psi| = |{\bf 10}^{\it 45}_1| = |\bar{\bf 5}^{\it 4}_1| = |\bar{\bf 5}^{\it5}_2| = |{\bf 1}_1|, 
\end{equation}
which are in the SM component fields,
\begin{equation}
|\psi| = |\nu^{c}_e| = |e| = |\nu_\mu| = |e^c|,
\end{equation}
and 
$\psi^4 = e e^c \nu_\mu \nu_e^c$.
In this direction, both baryon and lepton numbers are not generated by AD mechanism.




The $g$ coupling ($\phi \psi \psi^\prime$ term in the superpotential)
is needed to obtain the hybrid potential.
One can consider a variety of models for this coupling.
We will present examples to construct a model\footnote{
One can construct a model such as an SU(6) GUT
given in the original supernatural inflation paper \cite{Randall:1995dj}.
However, in their SU(6) example, there are unwanted fields,
and matter and unwanted fields are mixed intricately.
In the context of the flipped-SU(5), the model construction is much simpler,
and the quartic hilltop potential can be easily constructed.
}.

In the flipped-SU(5) model,
the right-handed neutrino is not a gauge singlet.
To acquire the right-handed neutrino Majorana mass,
the GUT symmetry breaking VEV is available.
For example,
one can consider the following superpotential to obtain the right-handed neutrino Majorana mass:
\begin{equation}
g \bar T {\bf 10} N + M_N N^2,
\label{eq46}
\end{equation}
where $N$ is a gauge singlet.
Then, the right-handed neutrino Majorana mass is obtained as
$g^2 \langle T \rangle^2/M_N$. 
More precisely, the seesaw mechanism to obtain the sub eV active neutrino mass
becomes so-called double seesaw.
Naively, $M_N$ is of the order of GUT scale (or Planck scale),
and the $g$ coupling should be much less than $O(1)$,
depending on the generation of ${\bf 10}$ we choose.
For example, if we choose $M_N$ to be the GUT scale (VEV of $T$),
suitable size of $g$ coupling is about 0.01
to obtain the right-handed Majorana mass to be $10^{12}$ GeV. 
Choosing the $D$-flat direction with the right-handed neutrino components (in {\bf 10}): e.g.,
\begin{equation}
|\psi| = |{\bf 10}^{\it 45}_2| = |{\bf 10}^{\it 12}_1| = |{\bf 10}^{\it23}_2| = |\bar{\bf 5}^{\it 2}_1|, 
\end{equation}
or
\begin{equation}
|\psi| = |{\bf 10}^{\it 45}_2| = |\bar{\bf 5}^{\it 4}_1| = |\bar{\bf 5}^{\it5}_2| = |{\bf 1}_1|, 
\end{equation}
the superpotential term ${\bf 10}_2 {\bf 10}_1 {\bf 10}_2 \bar{\bf 5}$
or ${\bf 10}_2 \bar{\bf 5}_1 \bar{\bf 5}_2 {\bf 1}_1$
can induce the hilltop potential.
From the $D$-flatness,
generations of ${\bf 10}^{\it 12}$ and ${\bf 10}^{\it23}$ have to be different.
This is same for the generations of 
$\bar{\bf 5}^{\it 4}$ and $\bar{\bf 5}^{\it5}$.


We note that the $O(1)$ size of the coupling $a$ (in the Planck mass unit) 
is favored for the quartic hilltop term
to obtain $n_s = 0.96$.
The ${\bf 10} \cdot \bar{\bf5} \cdot \bar{\bf5}  \cdot {\bf 1}$ 
and ${\bf 10} \cdot {\bf10} \cdot {\bf10} \cdot \bar{\bf 5}$ 
terms contain $d^c u^c c^c e^c$ and $qqq\ell$ terms, respectively.
The $O(1)$ size of this term can be problematic to cause a rapid proton decay
depending on the flavor configuration.
To satisfy the $F$-flatness, the flavor configuration has to be chosen (as described),
and the required hilltop term may contain the 1st and 2nd generation much.
Then, it is dangerous even if the sfermion masses are 100 TeV, especially for
the operator ${\bf 10} \cdot {\bf10} \cdot {\bf10} \cdot \bar{\bf 5}$,
which contains left-handed proton decay operator $qqq\ell$.
The gaugino dressing for the proton decay operators 
can suppress the proton decay amplitude
by a factor 
$m_{\rm gaugino}/m^2_{\tilde q}$.
The light gaugino scenario is favored to suppress the proton decay via the hilltop term.
We note that 
for the case of ${\bf 10} \cdot \bar{\bf5} \cdot \bar{\bf5}  \cdot {\bf 1} 
\supset \nu^c\ell\ell e^c + d^c u^c u^c e^c$,
one can choose ${\bf 1}_i = e^c_i$ to be the 3rd generation (right-handed tau lepton in the mass eigenstate),
but {\bf10} and $\bar{\bf5}$ are 1st and 2nd generations.
(In order not to lift the $F$-term potential from the Yukawa coupling, 
${\bf10}$ and $\bar{\bf5}$ are not given in the mass eigenstates for up-type quarks.
One can choose the $F$-flat flavor configuration for {\bf 10} and $\bar{\bf5}$, without loss of generality).
Then, the right-handed operator $d^c u^c u^c e^c$ is not dangerous because tau lepton is heavier than proton
unless a large flavor changing current is affected by SUSY breaking slepton masses.

One can adopt ${\bf 50} + \overline{\bf 50}$ representations
as a waterfall $\phi$ field,
whose VEV gives a right-handed neutrino Majorana mass directly
by a ${\bf 10}\cdot {\bf 10}\cdot \overline{\bf 50}$ coupling.
For example, there is a Majorana mass term
\begin{equation}
g {\bf 10}_2 {\bf 10}_3 \overline{\bf 50},
\end{equation}
which can be identified to the $g \phi \psi \psi^\prime$ coupling to obtain the hybrid potential,
for the right-handed neutrino component in ${\bf 10}_3$ is a (part of) inflaton,
and ${\bf 10}_2$ is identified as $\psi^\prime$ which does not contribute to the inflation.
The smallness of $g$ coupling can be related to the
Majorana mass scale,
which should be smaller than the GUT/Planck scale.
In this case, the VEV of ${\bf 50} + \overline{\bf 50}$ can also break the 
$SU(5)\times U(1)_X$ down to SM.
Contrary to the ${\bf 10} + \overline{\bf 10}$ waterfall fields,
they are not utilized to realize the doublet-triplet splitting.
Therefore, as a building block for this structure,
one can consider Pati-Salam model in a simple manner.

Before moving to describe the inflation in the Pati-Salam model,
we comment on the decays of inflaton and waterfall fields.
Sometimes, it is said that
the decay of the lightest particle of $\psi$, $\phi$ and $\psi^\prime$
is kinematically blocked
due to the symmetry to obtain the hybrid inflation potential.
Actually,
the superpotential term 
\begin{equation}
W = g \psi \phi \psi^\prime + \kappa S(\phi \bar \phi - \Lambda^2)
\end{equation}
has a parity symmetry
(assigning $-$ to $\psi^\prime$, $\phi$ and $\bar\phi$,
and $+$ to the inflaton $\psi$). 
Inflaton $\psi$ can decay to the SM particle, via the Yukawa coupling,
One may think that one of $\psi^\prime$ and $\phi$ cannot decay due to the kinematical block.
However,
the parity symmetry is violated via the non-zero values of the waterfall fields.
Let us explain it in a concrete example:
\begin{equation}
W = g_i \bar\phi \nu^c N_i + y \ell \nu^c H_u + (M_N)_{ij} N_i N_j,
\end{equation}
where the right-handed neutrino is the (part of) inflaton
and the waterfall field is $\bar \phi = \bar T^{45}$.
Because there has to be multiple $N$ fields to obtain the Majorana masses 
of three right-handed neutrinos,
we write the index of $N$ explicitly.
The inflaton field is the right-handed Majorana neutrino
and its mass is a little less than the unification scale, $g^2 \phi^2/M_N$
after falling down the waterfall potential.
The inflaton field can decay into MSSM field directly through the Dirac neutrino Yukawa coupling $y$.
The field $N$ is mixed with $\nu^c$ in $\bf 10$ via the non-zero value of $\phi$,
it can decay into the MSSM field directly (without decaying into $\phi$).
As a consequence, both $\phi$ and $\psi^\prime$ can decay into MSSM fields after all.
In the case where we adopt ${\bf 50}+\overline{\bf 50}$ as waterfall fields,
the situation is the same as above.


%

%
%
%




\section{Pati-Salam model}
\label{section5}

In the Pati-Salam model,
the gauge symmetry is $SU(4)_c \times SU(2)_L \times SU(2)_R$,
and the matter content of the model (See Appendix \ref{b} for more detail) is
\begin{equation}
\begin{array}{|c|c|}\hline
\mbox{Matter} &  SU(4)_c\times SU(2)_L \times SU(2)_R \\ \hline\hline
L_i  = (q_i,\ell_i) & ({\bf 4},{\bf 2},{\bf1}) \\ \hline
\bar{R}_i = (u^c_i,d^c_i, e^c_i, \nu^c_i)& (\bar{\bf 4},{\bf 1},{\bf2})  \\ \hline
\end{array}
\end{equation}

The Pati-Salam gauge symmetry is directly broken down to
SM gauge symmetry
by $({\bf 4},{\bf 1},{\bf 2}) + (\bar{\bf 4},{\bf 1},{\bf 2})$
or $({\bf 10},{\bf 1},{\bf 3}) + (\overline{\bf 10},{\bf 1},{\bf 3})$.
The non-zero values of them can generate the right-handed neutrino Majorana masses
as same as the flipped-SU(5) model.
Therefore, the hybrid inflation model with hilltop potential can be easily constructed.
Because the dimension of the representation is less than the one in the flipped-SU(5)
model, it is easier to understand the field contents in it as a building block.
In the model using $({\bf 10},{\bf 1},{\bf 3}) + (\overline{\bf 10},{\bf 1},{\bf 3})$
to obtain the waterfall potential,
the remnants in the representations can remain light
due to the accidental discrete symmetry for the waterfall potential.
The {\bf 10} dimensional representation in $SU(4)_c$ is a symmetric tensor,
and it can be decomposed under $SU(4)_c \to SU(3)_c \times U(1)_{B-L}$ to 
\begin{equation}
{\bf 10} = {\bf 6}_{2/3} + {\bf 3}_{-2/3} + {\bf 1}_{-2},
\end{equation}
where subscript denotes the $U(1)_{B-L}$ charge.
The $B-L$ charge can be obtained by
the $B-L$ generator,
\begin{equation}
T_{B-L} = {\rm diag} \left(\frac13,\frac13,\frac13,-1\right).
\end{equation}
The $SU(3)_c$ sextet can remain light till TeV scale,
and it can have an interesting phenomenological implication for
the baryon number violation \cite{Chacko:1998td}.
The implication in the inflation model will be studied in an accompanied paper.

The quartic invariants in terms of the matter representation to obtain the hilltop potential are
\begin{equation}
LLLL, \qquad
\bar R \bar R \bar R \bar R, \qquad
LL \bar R \bar R.
\end{equation}
Due to the $SU(4)_c$ symmetry, matter cubic terms are not allowed.
For $LLLL$, one can write the group indices explicitly,
\begin{equation}
\epsilon^{abcd} \epsilon_{\alpha\beta} \epsilon_{\gamma\delta} L_a^\alpha L_b^\beta L_c^\gamma L_d^\delta,
\end{equation}
where $\epsilon$ is a total anti-symmetric tensor, the indices $a,b,c,d$ are for $SU(4)_c$,
and $\alpha,\beta,\gamma,\delta$ are for $SU(2)_L$.
We do not write the group index explicitly later.
We note that all the generation of $L$ cannot be same
in the $LLLL$ operator due to anti-symmetricity.
If the $D$-flat direction to lift $LLLL$ and $\bar R\bar R \bar R\bar R$  potential,
the $F$-term potential from the Yukawa coupling is not lifted.
To use $LL\bar R \bar R$ potential,
one has to care about the generation configuration 
in order not to lift the $F$-term potential
as we have explained in the flipped-SU(5) model.

If $({\bf 4},{\bf 1},{\bf 2}) + (\bar{\bf 4},{\bf 1},{\bf 2})$ is employed to break the gauge symmetry,
$R$-parity violating terms can be generated since one of the fields has the same quantum number
as the right-handed field $\bar R$. 
If $({\bf 10},{\bf 1},{\bf 3}) + (\overline{\bf 10},{\bf 1},{\bf 3})$ is employed,
$R$-parity violating terms ($qd^c \ell, u^c d^c d^c, \ell \ell e^c, \ell H_u$ in MSSM) are not
generated after the gauge symmetry breaking.
(If the $B-L$ symmetry is broken by $B-L = \pm 2$ fields, the $R$-parity symmetry is preserved
because the $R$-parity corresponds to $Z_2$ subgroup of $U(1)_{B-L}$.)

In the Pati-Salam model, one can make the right-handed neutrino to be inflaton,
and to become heavy after the waterfall fields falls down,
similarly to the flipped-SU(5) model.
Here we describe a different configuration as an example.
The Higgs representation to break the elecroweak symmetry is bi-doublet representation:
\begin{equation}
H : ({\bf 1},{\bf 2},{\bf2}).
\end{equation}
The Yukawa interaction to generate masses of quarks and leptons is 
\begin{equation}
L \bar R H.
\end{equation}
If there is only one bi-doublet field,
all the up-type quark, down-type quark, and charged-lepton and Dirac neutrino Yukawa 
couplings are same.
Therefore, to break the wrong prediction of the fermion mass, one has to extend the model as
either (case 1) there are multiple bi-doublets, or (case 2) quarks and lepton fields are mixed with
the other representation.
The case 2 is compatible to the hybrid inflation scenario.

We exhibit the scenario to mix the right-handed strange quark field with the inflaton
by employing a field
\begin{equation}
S_D = ({\bf 6},{\bf 1},{\bf1}).
\end{equation}
The ${\bf6}$-dimensional representation is a anti-symmetric tensor of $SU(4)_c$.
The representation can be decomposed under $SU(3)_c \times SU(2)_L \times U(1)_Y$ as
\begin{equation}
D: ({\bf 3},{\bf 1},-1/3) + D^c: (\bar{\bf 3},{\bf 1},1/3).
\end{equation}
The representation is equivalent to the right-handed down-type quark's one,
and they can mix.
Denoting the fields to break the Pati-Salam symmetry as
\begin{equation}
\Phi: ({\bf 4},{\bf 1},{\bf 2}), \qquad \bar\Phi: (\bar{\bf 4},{\bf1},{\bf2}),
\end{equation}
we consider the following terms:
\begin{equation}
g S_D \bar R \bar\Phi + M_S S_D^2,
\end{equation}
where $g$ is a coupling constant.
After the $\bar\Phi$ field acquire a VEV $\bar\phi$, the
fields $D^c$ and $d^c$ in $\bar R$ are mixed:
\begin{equation}
g D d^c \bar\phi + M_S D D^c = D (g d^c \bar \phi + M_S D^c),
\end{equation}
and a linear combination of them has a mass $\sqrt{g^2 \bar\phi^2 + M_S^2}$,
while another linear combination, ($M_S d^c - g \bar\phi D^c$) remains massless
(up to the electroweak scale VEV by the bi-doublet $H$).
Through this kind of interaction,
the quarks and leptons are mixed
and the unification of fermion mass can be violated
to obtain realistic fermion masses in SM
even if there is only one bi-doublet Higgs.
The other candidates to mix with the MSSM matter are the followings:
\begin{eqnarray}
({\bf 6},{\bf 2},{\bf 2}) &=& ({\bf 3},{\bf 2},1/6) + ({\bf 3},{\bf 2},-5/6) + 
(\bar{\bf 3},{\bf 2},5/6) + (\bar{\bf 3},{\bf 2},-1/6),  \\
({\bf 6},{\bf 1},{\bf 3}) &=& ({\bf 3},{\bf 1},2/3) + ({\bf 3},{\bf 1},-1/3) + ({\bf 3},{\bf 1},-4/3) +
(\bar{\bf 3},{\bf 1},4/3) + (\bar{\bf 3},{\bf 1},1/3)  + (\bar{\bf 3},{\bf 1},-2/3), \\
({\bf 1},{\bf2},{\bf2}) &=& ({\bf 1},{\bf2},1/2) + ({\bf1},{\bf2},-1/2), \\
({\bf1},{\bf1},{\bf3}) &=& ({\bf1},{\bf1},1)+({\bf1},{\bf1},0)+({\bf1},{\bf1},-1).
\end{eqnarray}
They have components which can mix with $q$, $u^c$, $\ell$, $e^c$
respectively.

In the Pati-Salam model 
and left-right gauge model ($SU(3)_c \times SU(2)_L \times SU(2)_R \times U(1)_{B-L}$),
the $g D d^c \bar\phi$ term is motivated.
The non-zero value of $\bar\phi$ breaks the gauge symmetry down to SM one.
It is interesting that it is compatible with the hybrid inflation model.
For example, we can consider the following $D$-flat direction:
\begin{equation}
|\psi| = |t^c| = |c^c|= |s^c|=|\tau^c|.
\end{equation}
One can easily check that this satisfies the $D$-flatness condition for 
$SU(4)_c\times SU(2)_R$.
The $s^c$ field is mixed with the $D^c$ field in $S_D$ by $g s^c D \bar\phi$ term,
which generates the hybrid potential.
The hilltop term is obtained by
\begin{equation}
W = \frac{a}{M_P} \bar R_3 \bar R_2 \bar R_2 \bar R_3,
\end{equation}
where the subscripts denotes the generation index.
We note that 
this operator contains
\begin{equation}
t^c b^c s^c \nu_\mu^c + c^c b^c s^c \nu^c_\tau + t^c c^c s^c \tau^c + t^c c^c b^c \mu^c,
\end{equation}
and
the terms do not cause a problem of a rapid proton decay directly (unless
the right-handed top is mixed to the other generation sizably by $({\bf 6},{\bf1},{\bf3})$
or via gluino flavor change dressing),
because a suitable generation can be chosen 
with satisfying the $F$-flatness,
contrary to the case of flipped-SU(5) model.

In this simplified description,
the inflaton which oscillates after the inflation and generates entropy via its decay
is a linear combination,
$\hat\psi = s^c\cos\theta  +  D^c\sin\theta$,
and the
light strange quark is 
$\hat s^c = -s^c\sin\theta  + D^c\cos\theta $,
where $\tan\theta = M_s/(g\bar\phi)$.
The essential superpotential terms are
\begin{equation}
y q s^c H + g s^c D \bar \phi + M_S D D^c + \frac{a}{M_P} t^c c^c s^c \tau^c.
\end{equation}
Since the inflaton is mixed with the matter by the non-zero value of the waterfall field $\phi$,
it can decay to MSSM field via the Yukawa interaction term, $q s^c H = \cos \theta q \hat\psi H -\sin\theta q \hat s^c H $.
The $\psi^\prime$ field corresponds to $D$.
The Lagrangian includes a term
\begin{equation}
-{\cal L} \supset \left|\frac{\partial W}{\partial s^c}\right|^2
= \left| y q H + g D \bar \phi \right|^2 = g D \langle\bar\phi\rangle (y q H)^* + \cdots.
\end{equation}
Therefore, after $\bar\phi$ acquires a non-zero value, $\psi^\prime$ can directly decay into MSSM fields.
In this way, the decays are not kinematically blocked after all.

\section{Topological defects}
\label{section6}

When a gauge symmetry $G$ breaks down to $H$, 
monopoles are produced if the second homotopy group $\pi_2(G/H)$ is non-trivial
\cite{'tHooft:1974qc}.
In fact, the standard model gauge group contains $U(1)_Y$,
and therefore, it is possible that the monopoles are produced if the
unified gauge group is semi-simple 
(e.g. $SU(5)$, $SO(10)$, and $SU(4)_c \times SU(2)_L \times SU(2)_R$). 
As it is well-known, the inflation (after the GUT symmetry breaking phase transition) can 
be motivated to dilute the monopole density.

In the current setup, the phase transition by waterfall potential occurs after the inflation ends,
and the vacuum manifold has to be considered to avoid the monopole problem.
Because the $\psi$ fields have quantum numbers of the unified gauge group,
the unified gauge symmetries 
(i.e. $SU(5) \times U(1)_X$ and $SU(4)_c \times SU(2)_L \times SU(2)_R$ in our models)
are not fully maintained during inflation.
Due to the situation, the monopoles are not necessarily generated at the
waterfall phase transition even in the case of Pati-Salam semi-simple gauge group,
depending on the $D$-flat configuration of $\psi$ fields.

Let us investigate 
how the configuration of 
$\bar R: (\bar{\bf4},{\bf1},{\bf2})$ along the $D$-flat direction
breaks the $SU(4)_c \times SU(2)_R$ gauge group
in the Pati-Salam model.
As given in Appendix \ref{b}, the representation contains the 
the right-handed matter fields as
\begin{equation}
\bar R_i^\alpha = \left(
\begin{array}{cccc}
  u^c_r & u^c_g & u^c_b & \nu^c \\
  d^c_r & d^c_g & d^c_b & e^c
\end{array}
  \right),
\end{equation}
where sub(super)script of $\bar R$ stands for the $SU(4)_c$ ($SU(2)_R$) index.
As the first example, let us consider the $D$-flat configuration in terms of 2nd and 3rd generations:
\begin{equation}
(\bar R_2)_i^\alpha = \left(
\begin{array}{cccc}
 a & 0 & 0 & 0 \\
 0 & b & 0 & 0 
\end{array}
  \right),
  \qquad
  (\bar R_3)_i^\alpha = \left(
\begin{array}{cccc}
 0 & 0 & 0 & d \\
 0 & 0 & c & 0 
\end{array}
  \right),
\end{equation}
which corresponds to $\psi^4 = c^c s^c b^c \nu_\tau^c \neq 0$.
The $D$-flat conditions for $SU(4)_c$ and $SU(2)_R$ 
($\sum_{I=2,3} \sum_\alpha (\bar R_I)_i^\alpha (\bar R_I)_j^\alpha \propto \delta_{ij}$
and
$\sum_{I=2,3} \sum_i (\bar R_I)_i^\alpha (\bar R_I)_i^\beta \propto \delta_{\alpha\beta}$)
are satisfied if $|a|= |b|=|c|=|d|$.
Along the $D$-flat direction, one can find that the remained symmetry is $SU(2)$,
which is a linear combination of $SU(2)\times SU(2) \subset SU(4)_c$ and $SU(2)_R$.

The remained $SU(2)$ symmetry can be broken at the phase transition by the waterfall fields
$\Phi$ and $\bar\Phi$.
Suppose that the waterfall fields are 
\begin{equation}
\Phi : ({\bf 4},{\bf1},{\bf2}), \qquad \bar\Phi : (\bar{\bf4}, {\bf1},{\bf2}),
\end{equation}
and the vacuum configuration is
\begin{equation}
\Phi = 
\left(\begin{array}{cccc}
  0 & 0 & 0 & \phi \\
  0 & 0 & 0 & 0
\end{array}
  \right),
 \quad
\bar \Phi = 
\left(\begin{array}{cccc}
  0 & 0 & 0 & \bar\phi \\
  0 & 0 & 0 & 0
\end{array}
  \right). 
  \label{phi-conf}
\end{equation}
In this case, the remained $SU(2)$ symmetry is broken down to $U(1)$.
As a result, at the phase transition (after the inflation ends), the monopoles are generated.

The vacuum configuration of the waterfall fields can be chosen to be
Eq.(\ref{phi-conf}) without loss of generality.
On the other hand, the $D$-flat configuration of $\psi$ (which causes the inflation) is not necessarily 
aligned to the $\Phi$ direction.
For example, 
one can consider the configuration,
\begin{equation}
 (\bar R_2)_i^\alpha = U \left(
\begin{array}{cccc}
  a & 0 & 0 & 0 \\
  0 & b & 0 & 0
\end{array}
  \right),
  \qquad
   (\bar R_3)_i^\alpha = U \left(
\begin{array}{cccc}
  0 & 0 & 0 & d \\
  0 & 0 & c & 0
\end{array}
  \right),
\end{equation}
where $U$ is a $2\times2$ unitary matrix.
The $D$-flat conditions can be satisfied similarly if $|a|=|b|=|c|=|d|$,
and $SU(2)$ symmetry is remained along the $D$-flat direction.
At the phase transition by the waterfall field, the $SU(2)$ symmetry is completely broken
unless $U$ is a diagonal matrix.
Therefore, the monopoles are not necessarily generated at the phase transition.
Only when the $\psi$ and $\phi$ configuration are aligned, the monopoles can be generated.

Next, let us consider the $D$-flat configuration in terms of three generations.
For example,
\begin{equation}
 (\bar R_1)_i^\alpha = \left(
\begin{array}{cccc}
  a & 0 & 0 & 0 \\
  0 & b & 0 & 0
\end{array}
  \right),
  \qquad
   (\bar R_2)_i^\alpha = \left(
\begin{array}{cccc}
  0 & 0 & 0 & 0 \\
  0 & 0 & c & 0
\end{array}
  \right),
  \qquad
     (\bar R_3)_i^\alpha = \left(
\begin{array}{cccc}
  0 & 0 & 0 & d \\
  0 & 0 & 0 & 0
\end{array}
  \right).
\end{equation}
The $D$-flat conditions are satisfied if $|a|= |b|=|c|=|d|$, again.
In this case, one can find that the remained symmetry along the flat direction is $U(1)$,
and the remained $U(1)$ symmetry is not broken at the waterfall phase transition.
Therefore, no topological defects are generated in this example.
Similar to the two generation case, it is not necessary to align the $\psi$ and $\phi$ 
configuration, and a unitary matrix can be multiplied.
In general, therefore, the remained $U(1)$ symmetry is broken at the phase transition,
and a cosmic string can be generated.
However, the broken $U(1)$ symmetry is recovered after the $\psi$ field settles on the vacua,
and the cosmic string disappear.

We stress that the monopole production can be avoided at the waterfall phase transition
 in the current setup of the Pati-Salam model.
We also note that in the flipped-SU(5) model, no serious topological defects are generated at the
waterfall phase transition.
For example, in the case of $D$-flat configuration in Eq.(\ref{Dflat-fSU5}),
the remained symmetry is $SU(3) \subset SU(5)$ and it breaks down to $SU(2)$ by the waterfall fields.

\section{Non-thermal Leptogenesis and gravitino problem}
\label{section7}

The inflaton in our model is a flat direction containing right-handed sneutrino. It has a soft mass roughly $10$ TeV during inflation. However, after inflation, the GUT Higgs develops a VEV and the right-handed sneutrino becomes massive by the double seesaw mechanism\footnote{A right-handed sneutrino field which acquires a large mass via coupling to the waterfall field was also considered in \cite{Berezhiani:2001xx} where different from our model, Affleck-Dine leptogenesis was considered and right-handed sneutrino is not the inflaton.}. It decays mainly via Yukawa coupling $y$ into slepton and Higgs or into lepton and Higgsino with a decay width given by $\Gamma_N=M_N y^2 / (4 \pi)$. The decay of the sneutrino after inflation reheats the universe to a temperature $T_R \sim \sqrt{\Gamma_N M_P}$ if the waterfall field decays earlier enough which we will assume to be the case for a simple estimation. In this case, non-thermal leptogenesis may happen. The baryon asymmetry is given by
\begin{equation}
\frac{n_B}{n_\gamma} \sim \varepsilon \frac{T_R}{M_N} \sim 10^{-10} \left( \frac{T_R}{10^6\mbox{ GeV}} \right) \delta,
\end{equation}
where $\varepsilon \sim \frac{3}{8 \pi}\sqrt{\Delta m^2_{31}} M_N \delta/ (\langle H_u \rangle ^2)$, $\Delta m^2_{31} \sim 2.6 \times 10^{-3} \mbox{eV}^2$ is the atmospheric neutrino mass squared difference and $\langle H_u \rangle \sim 174\mbox{ GeV}$, and $\delta$ is the effective CP violating phase \cite{Murayama:1992ua, Hamaguchi:2001gw}. Therefore successful non-thermal leptogenesis can happen if the temperature at right-handed sneutrino decay is $\gtrsim 10^6$ GeV \cite{Antusch:2004hd, Kadota:2005mt, Antusch:2010mv}. 

Since the gravitino mass we consider is $100$ TeV, the gravitino does not affect big bang nucleosynthesis (BBN).
However, an upper bound of reheating temperature $\sim 10^{10}\mbox{ GeV}  (100\mbox{ GeV}/m_{\rm LSP})$, with $m_{\rm LSP}$ LSP mass is given by the production of LSP cold dark matter. Interestingly we may have both leptogenesis and dark matter if this upper bound is saturated.


\section{Conclusion and Discussion}
\label{con}

In this paper, we present two concrete models of hilltop supernatural inflation based on flipped SU(5) and Pati-Salam models. The phenomenology in our model is very rich. Our inflation model is closely connected to particle physics. For example, our parameter space is constrained by both proton decay and CMB. Hilltop inflation fits very well in recent PLANCK data concerning spectral index, non-Gaussianity, tensor to scalar ratio and basically all the observables.  As a hybrid inflation model, we consider a flat direction containing right-handed sneutrino to be the inflaton field and the waterfall field is a GUT Higgs. Non-thermal leptogenesis can happen after inflation. It is also possible to generate LSP dark matter. We have also shown that topological defects are not produced in the current setups.

\section*{Acknowledgement}
This work is partially supported by the Grant-in-Aid for Scientific
research from the Ministry of Education, Science, Sports, and Culture, Japan, Nos.
21111006, 22244030, 23540327 (K.K.), 21244036, 23654090, 23104009 (C.S.L), and 21244036 (C.M.L.).
The work of Y.M. is supported by the Excellent Research Projects of National Taiwan University under grant number NTU-98R0526.


\appendix

\section{inflation calculation}
\label{a}

We consider a potential of the hilltop form,
\begin{equation}
V(\psi)= V_0 \left( 1+\frac{1}{2}\eta_0 \frac{\psi^2}{M_P^2}\right)-\lambda \frac{\psi^p}{M_P^{p-4}},
\end{equation}
where $V_0/M_P^4= 10^{-24}$ in our model.

The number of e-folds is given by
\begin{equation}
N=M^{-2}_P\int^{\psi}_{\psi_{end}}\frac{V}{V'}d\psi.
\label{efolds}
\end{equation}

This model can be solved analytically and we have

\begin{eqnarray}
\left(\frac{\psi}{M_P}\right)^{p-2}&=&\left(\frac{V_0}{M_P^4}\right)
\frac{\eta_0 e^{(p-2)N\eta_0}}{\eta_0 x+4 \lambda (e^{(p-2)N\eta_0}-1)} \label{a2}\\
x &
\equiv & \left(\frac{V_0}{M_P^4}\right) \left(\frac{M_P}{\psi_{end}}\right)^{p-2}, \label{a3}
\end{eqnarray}

The spectrum and the spectral index are given respectively by
\begin{eqnarray}
P_R&=&\frac{1}{12\pi^2} \left(\frac{V_0}{M_P^4}\right)^{\frac{p-4}{p-2}} e^{-2N\eta_0}\frac{[p\lambda(e^{(p-2)N\eta_0}-1)+\eta_0 x]^{\frac{2p-2}{p-2}}}{\eta_0^{\frac{2p-2}{p-2}}(\eta_0 x-p\lambda)^2} \label{a4}\\
n_s&=&1+2\eta_0 \left[1-\frac{\lambda p(p-1) e^{(p-2)N\eta_0}}{\eta_0 x+p\lambda(e^{(p-2)N\eta_0}-1)}\right]. \label{a5}
\end{eqnarray}
From the above equations we can obtain 
\begin{equation}
\lambda=\frac{(12 \pi^2 P_R)^{\frac{p-2}{2}}}{p[2(p-1)]^{(p-1)}} 
\left( \frac{V_0}{M_P^4} \right)^{-\frac{p-4}{2}}(2\eta_0+1-n_s)(2(p-2)\eta_0-1+n_s)^{(p-2)}.
\end{equation}

\subsection{$p=4$}

\begin{figure}[t]
  \centering
\includegraphics[width=0.6\textwidth]{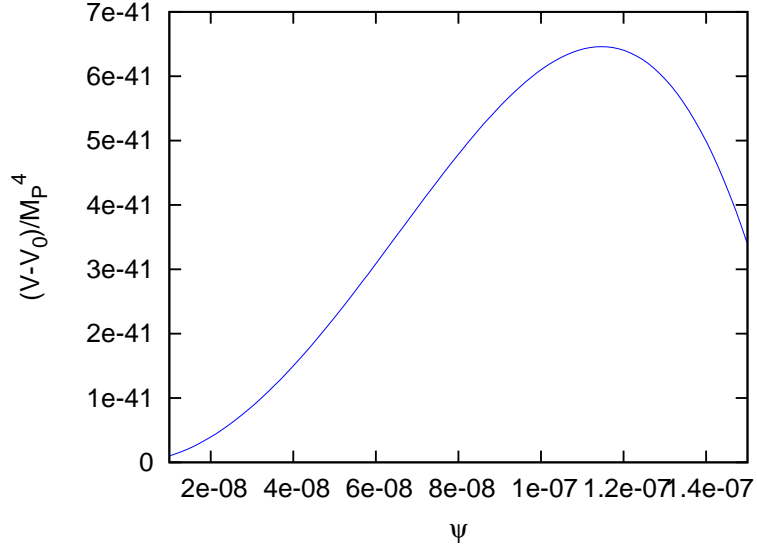}
  \caption{$(V-V_0)/M_P^4$ as a function of $\psi$. This plot shows the true shape of our potential which is made by including the $\psi^6$ term. The hilltop form is clear.}
  \label{potentialform}
\end{figure}

\begin{figure}[t]
  \centering
\includegraphics[width=0.6\textwidth]{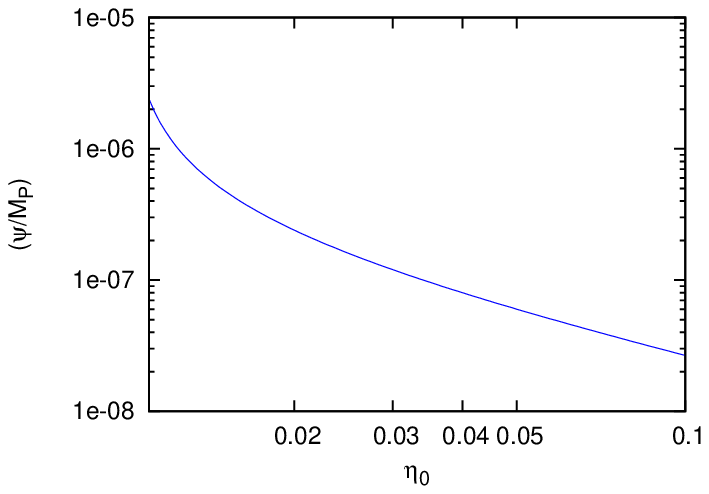}
  \caption{$\psi/M_P$ as a function of $\eta_0$ for $p=4$. For $\eta_0=0.02$, we have $\psi \sim 10^{-7} M_P$.}
  \label{fig2}
\end{figure}

\begin{figure}[t]
  \centering
\includegraphics[width=0.6\textwidth]{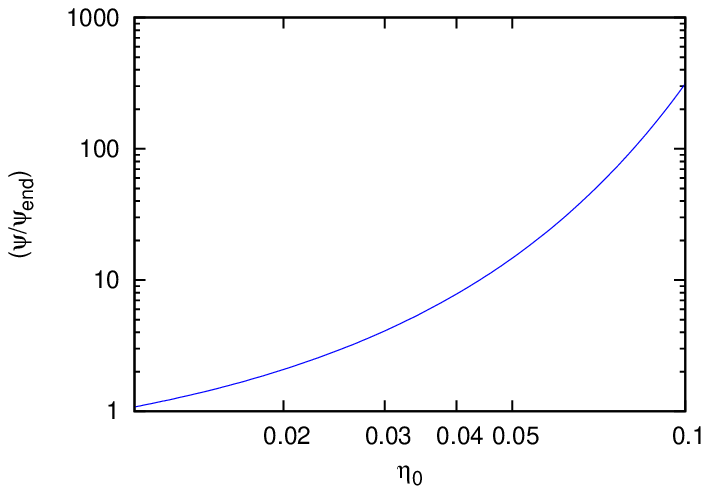}
  \caption{$\psi/\psi_{end}$ as a function of $\eta_0$ for $p=4$. In particular, for $\eta_0=0.02$, $\psi$ and $\psi_{end}$ are the same order.}
  \label{fig3}
\end{figure}

Here $\psi$ means the field value of inflaton at number of e-folds $N$ and $\psi_{end}$ means the field value of inflaton at $N=0$. By imposing the CMB normalization $P_\zeta^{1/2} = 5 \times 10^{-5}$ and $n_s=0.96$ at $N=60$, we can solve $x$ from Eq.~(\ref{a5}) as a function of $\lambda$ and $\eta_0$. Then substitute $x$ into Eq.~(\ref{a4}), we can obtain the relation between $\eta_0$ and $\lambda$ given by
\begin{equation}
\lambda=1.1\times 10^{-8} \times (\eta_0-0.01)^2 (\eta_0+0.02).
\label{lambda}
\end{equation}
This is ploted in Fig.~\ref{fig1}. For example, for $\eta_0=0.04$, we have $\lambda=5.9 \times 10^{-13}$. By using those numbers, we show the typical potential form in Fig.~\ref{potentialform}.

Having the relation of $\lambda$ as a function of $\eta_0$, we can express $\psi$ (at $N=60$) as a function of a single parameter $\eta_0$ by using Eq.~(\ref{a2}) and obtain 
\begin{equation}
\frac{\psi}{M_P} =\frac{2.4 \times 10^{-9}}{\eta_0-0.01}.
\end{equation}
This is ploted in Fig.~\ref{fig2}. In order to know $\psi_{end}$, we can use Eqs.~(\ref{a2}) and (\ref{a3}) to show
\begin{equation}
\frac{\psi}{\psi_{end}}  = \left( \frac{(2\eta_0-0.02)e^{120\eta_0}+(\eta_0+0.02)}{3\eta_0} \right)^{1/2}.
\end{equation}
This is ploted in Fig.~\ref{fig3}.

\subsection{$p=6$}

\begin{figure}[t]
  \centering
\includegraphics[width=0.6\textwidth]{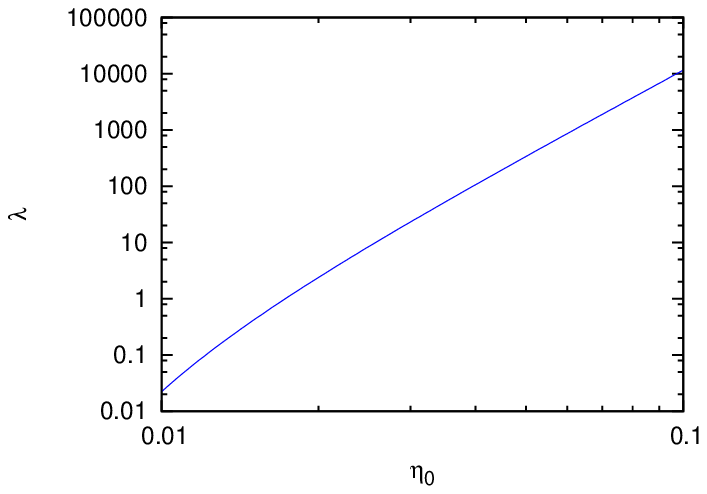}
  \caption{$\lambda$ as a function of $\eta_0$ for $p=6$. This plot is made by imposing $n_s=0.96$ and $P_\zeta^{1/2}=5 \times 10^{-5}$.}
  \label{fig5}
\end{figure}

By the same calculation procedure as $p=4$ case,
the relation between $\lambda$ and $\eta_0$ is given by
\begin{equation}
\lambda = 7.4 \times 10^7 \times (2\eta_0-0.01)^4(0.02+\eta_0).
\end{equation}
This is plotted in Fig.~\ref{fig5}.

\begin{figure}[t]
  \centering
\includegraphics[width=0.6\textwidth]{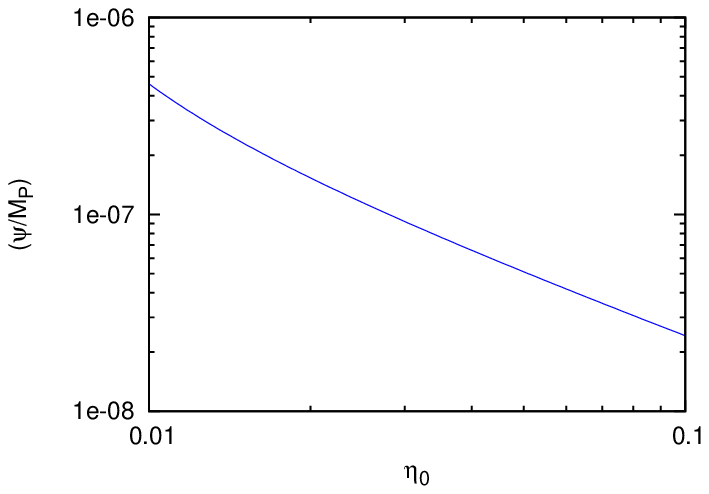}
  \caption{$\psi/M_P$ as a function of $\eta_0$ for $p=6$. For $\eta_0=0.02$, we have $\psi \sim 10^{-7} M_P$.}
  \label{fig6}
\end{figure}

\begin{figure}[t]
  \centering
\includegraphics[width=0.6\textwidth]{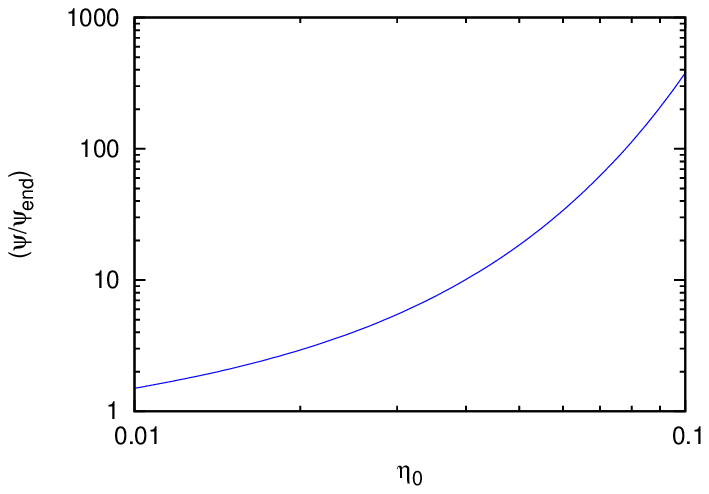}
  \caption{$(\psi/\psi_{end})$ as a function of $\eta_0$ for $p=6$. In particular, for $\eta_0=0.02$, $\psi$ and $\psi_{end}$ are the same order.}
  \label{fig7}
\end{figure}

The field value at $N=60$ is given by
\begin{equation}
\frac{\psi}{M_P} =\frac{4.6 \times 10^{-9}}{2\eta_0-0.01}.
\end{equation}
This is plotted in Fig.~\ref{fig6}.
In order to know $\psi_{end}$ we can obtain
\begin{equation}
\frac{\psi}{\psi_{end}}  = 0.67 \left( \frac{(4\eta_0-0.02)e^{240\eta_0}+(\eta_0+0.02)}{\eta_0} \right)^{1/4}.
\end{equation}
This is plotted in Fig.~\ref{fig7}.

\section{Unified Models}
\label{b}

\subsection{Flipped-SU(5)}

The SO(10) symmetry has a maximal subgroup $SU(5) \times U(1)$.
There are two different hypercharge assignment under the symmetry.
The SU(5) symmetry has a maximal subgroup $SU(3) \times SU(2) \times U(1)_{Y^\prime}$,
and the generator for the U(1) subgroup is
\begin{equation}
T_{Y^\prime} = {\rm diag} \left(\frac13,\frac13,\frac13,-\frac12,-\frac12\right).
\end{equation}
In the usual SU(5) GUT model,
the hypercharge $Y$ is same as $Y^\prime$ ($T_Y = T_{Y^\prime}$).
Therefore, the matter contents in the usual SU(5) are
\begin{equation}
{\bf 10}_{ab} = \left(
\begin{array}{ccccc}
0 & u^c_b & - u^c_g & u_r & d_r \\
-u^c_b & 0 & u^c_r & u_g & d_g \\
u^c_g & - u^c_r & 0 & u_b & d_b \\
-u_r & -u_g & - u_b & 0 & e^c \\
-d_r & - d_g & - d_b & - e^c & 0
\end{array}
\right),
\end{equation}
\begin{equation}
\bar{\bf 5}^a = (d^c_r, d^c_g, d^c_b, e, -\nu),
\end{equation}
where $r,g,b$ are color indices, $a,b$ are the SU(5) group indices, and $c$ stands for anti-reps.
The hypercharge of $\bf 10$ representation
is obtained by $(T_{Y})_a^{a^\prime} {\bf 10}_{a^\prime b} +   (T_{Y})^{a^\prime}_b {\bf 10}_{a a^\prime}$,
and $-(T_{Y}^*)_{a^\prime}^a \bar{\bf 5}^{a^\prime}$ for $\bar{\bf 5}$ representation.
The right-handed neutrino $\nu^c$ is singlet under the usual SU(5).

Under the $SU(5)\times U(1)_X$,
the $X$ charges are assigned to the $SU(5)$ representations
\begin{equation}
({\bf 10},1), \qquad (\bar{\bf5}, -3), \qquad  ({\bf 1}, 5).
\end{equation}
It is easy to check that the gauge anomaly is absent.
This is obvious because the $SU(5) \times U(1)_X$ is a subgroup of SO(10),
and the above matter multiplets can be embedded into the $\bf 16$ representation
under SO(10).
In the flipped-SU(5) model, the hypercharge in the standard model
is assigned as
\begin{equation}
Y = \frac15 (X - Y^\prime).
\end{equation}
It is easy to find that 
the places of the right-handed quarks ($u^c$ and $d^c$) and leptons ($\nu^c$ and $e^c$) are flipped,
respectively,
compared to the usual SU(5) assignment.

\begin{equation}
{\bf 10}_{ab} = \left(
\begin{array}{ccccc}
0 & d^c_b & - d^c_g & u_r & d_r \\
-d^c_b & 0 & d^c_r & u_g & d_g \\
d^c_g & - d^c_r & 0 & u_b & d_b \\
-u_r & -u_g & - u_b & 0 & \nu^c \\
-d_r & - d_g & - d_b & - \nu^c & 0
\end{array}
\right),
\end{equation}
\begin{equation}
\bar{\bf 5}^a = (u^c_r, u^c_g, u^c_b, e, -\nu),
\qquad
{\bf1} = e^c.
\end{equation}

The Higgs multiplets which include MSSM Higgs doublets ($H_u, H_d$) are:
\begin{eqnarray}
H : ({\bf 5}, -2),
\quad
\bar H : (\bar{\bf 5}, 2).
\end{eqnarray}
The component can be written as
\begin{equation}
H_a  = (H_C^r, H_C^g, H_C^b, H_d^0, H_d^-),
\quad
\bar H^a = (\bar H_C^r, \bar H_C^g, \bar H_C^b, H_u^0, -H_u^+).
\end{equation}

In the flipped-SU(5) model, the gauge symmetry is not a simple group,
and one may say that it is not a complete unified theory.
In the flipped-SU(5) model, however,
there is no gauge singlet in the SM matter with the right-handed neutrino,
and the model can be predictive.
As it is written in the text,
the doublet-triplet splitting is easily realized
by so-called missing partner mechanism 
with a single VEV of the field
$T: ({\bf 10}, 1)$ and $\bar T : (\overline{\bf 10},-1)$.
The 45 component of them (which is same as the right-handed neutrino components in the matter reps)
are singlets under the SM gauge symmetry,
and the flipped-SU(5) gauge symmetry is broken down to SM gauge symmetry
by those VEVs.

\subsection{Pati-Salam model}

The SO(10) symmetry has a maximal subgroup 
$SO(6) \times SO(4) \simeq SU(4) \times SU(2) \times SU(2)$.
The model with the gauge symmetry $SU(4)_c \times SU(2)_L \times SU(2)_R$
is called Pati-Salam model.
The $SU(4)_c$ symmetry is broken down to $SU(3)_c \times U(1)_{B-L}$,
and the hypercharge $U(1)_Y$ is a linear combination of $U(1)$ subgroup of $SU(2)_R$
and $U(1)_{B-L}$:
\begin{equation}
Y = \frac{B-L}{2} + T_{SU(2)_R}^3.
\end{equation}
The matter fields in SM are embedded in $({\bf 4},{\bf2},{\bf1})$ and $(\bar{\bf4},{\bf1},{\bf2})$:
\begin{equation}
({\bf 4},{\bf2},{\bf1}) = 
\left(
\begin{array}{cccc}
u_r & u_g & u_b & \nu \\
d_r & d_g & d_b & e
\end{array}
\right),
\qquad
(\bar{\bf 4},{\bf1},{\bf2}) = 
\left(
\begin{array}{cccc}
u_r ^c& u_g^c & u_b^c & \nu^c \\
d_r^c & d_g^c & d_b^c & e^c
\end{array}
\right).
\end{equation}
The MSSM Higgs doublets $(H_u,H_d)$ are embedded in bi-doublet representation,
\begin{equation}
({\bf1},{\bf2},{\bf2})=
\left(
\begin{array}{cc}
H_u^+& H_d^0 \\
H_u^0 & H_d^-
\end{array}
\right).
\end{equation}
The Pati-Salam gauge symmetry does not include abelian symmetry,
and the hypercharge is quantized,
which is one of the conceptual motivation of the unified models.

\end{document}